\documentclass[a4paper,11pt]{article}
\pdfoutput=1 % if your are submitting a pdflatex (i.e. if you have
\usepackage{jcappub} % for details on the use of the package, please
\usepackage[T1]{fontenc} % if needed

\newcommand{\hmpc}{{\, h^{-1}\, {\rm Mpc}}}

\title{\boldmath Can a conditioning on stellar mass explain the
  mutual information between morphology and environment?}

\author[a,1]{Snehasish Bhattacharjee,\note{Corresponding author.}}
\author[b]{Biswajit Pandey,}
\author[b]{Suman Sarkar}
\affiliation[a]{Department of Astronomy, Osmania University,
  Hyderabad, 500007, India} \affiliation[b]{Department of Physics,
  Visva-Bharati University, Santiniketan, 731235, India}
\emailAdd{snehasish.bhattacharjee.666@gmail.com}
\emailAdd{biswap@visva-bharati.ac.in}
\emailAdd{suman2reach@gmail.com}
\abstract{ Recent studies with SDSS have shown that a statistically
  significant non-zero mutual information between morphology and
  environment persists up to several tens of Mpc, which
    awaits an explanation. Galaxies in different environments acquire
  their stellar mass through accretion and merger and the stellar mass
  function of galaxies is known to depend on both environment and
  morphology. Naturally, stellar mass can be an important link between
  morphology and environment which may explain the non-zero mutual
  information between the two. Measuring the mutual information
  between morphology and environment by conditioning the stellar mass
  would allow us to test this possibility. We employ
    here a volume and stellar mass limited sample from the $16^{th}$
  data release (DR16) of the SDSS and find a non-zero
  conditional mutual information throughout the entire length scales
  probed. We compare the results with three different semi-analytic
  models implemented on the Millennium simulation and find their
  predictions to be in fairly good agreement with SDSS on smaller
  length scales ($\lesssim 30 \hmpc$), with a clear
    discrepancy observed at larger length scales ($\gtrsim 30 \hmpc$)
    where the models predict significantly lower conditional mutual
    information than the SDSS. Our analysis therefore suggests that
  only environmental and morphology dependence of
    stellar mass are inadequate in explaining the observed mutual
    information between morphology and environment and that physical
  processes which alters morphology may not necessarily have an impact
  on the stellar mass of galaxies and vice versa. }

\begin{document}
\maketitle
\flushbottom

\section{Introduction}
Understanding the formation and evolution of galaxies remains one of
major goals of modern cosmology. Environment is known to play a
governing role in the formation and evolution of galaxies. It is now
well known that environment imparts a significant influence on
different galaxy properties \cite{davis,guzzo,zehavi2002,goto,hogg,blanton,einasto,kauffmann,mouhcine,koyama}.
The massive ellipticals are found to reside in rich clusters and
spirals predominantly belong to groups or fields
\cite{hubble,zwicky,oemler,dressler}. This fact was further
re-established using the two point correlation function in
\cite{willmer,webstar,zehavi} where ellipticals were found to be
clustered richly in contrast to spirals. In a study of the
filamentarity of the galaxy distribution \cite{pandey2006}, spirals
were found to be distributed sporadically in the filaments whereas
ellipticals favored to reside densely in the nodes of the filament
intersections.

The local density is the most widely used indicator for the
environment of a galaxy. But the local density alone may not solely
describe the environmental dependence of galaxy properties.  The dark
matter haloes assemble in a hierarchical fashion and the galaxies are
believed to form later inside these haloes via radiative cooling and
condensation of baryons \cite{white}. In the halo model, the mass of a
dark matter halo decides all the properties of the galaxy formed
within it \citep{cooray}. However the different assembly history of
the dark matter haloes across the different parts of the cosmic web
leads to different clustering for them which is known as the assembly
bias \citep{croton, gao, musso, vakili}.  The galaxies in the Universe
are distributed along a vast interconnected network of sheets,
filaments and clusters encompassing nearly empty regions in between
them. The highly anisotropic nature of these structures lead to
variations in mass, shape, size and spin of dark matter haloes
residing in these large-scale environments \cite{hahn1}. So the
large-scale cosmic environment may also play a significant role in the
galaxy formation and evolution.

A number of observations indicate that there are significant
correlation between the galaxy properties and their large-scale
environment.  Employing data from SDSS, Scudder \cite{scudder} report
substantial difference in the star formation rates between groups
embedded in high density superclusters and the groups residing in
isolation. Using data from SDSS, Luparello \cite{luparello} report
strong dependence between properties of late type brightest group
galaxies and their large scale environment. An analysis of the
filamentarity of the galaxy distribution using SDSS by Pandey \&
Bharadwaj \cite{pandey08} reveal that average filamentarity of the red
galaxies are significantly lower than the starburst galaxies. Darvish
\cite{darvish} report that the presence of filamentary environment
increases the fraction of starburst galaxies at $z \sim 1$. An
analysis of the galaxies in SDSS by Filho \cite{filho} reveal that
almost three quarters of the most extremely metal poor galaxies are
embedded in voids and sheets. A recent work by Pandey \& Sarkar
\cite{pandey20} find that the fraction of red galaxies in filaments
and sheets increases with the size of these structures. Park
\cite{park} analyzed galaxies from SDSS and find weak dependency
between large scale environment and galaxy properties once the
contributions of morphology and luminosity are accounted for. Yan
\cite{yan} employed data from SDSS and report that the tidal
environment of the large scale structures has no influence on the
galaxy properties. \\Pandey \& Sarkar \cite{pandey} analyzed the data
from SDSS and report that morphology and environment share a non-zero
mutual information (hereafter MI) which decreases with increasing length scales. A
recent analysis by Sarkar \& Pandey \cite{sarkar} show that these
MI are statistically significant at $99.9\%$
confidence level throughout the entire length scale probed. In the
present work, we would like to understand the physical origin of such
non-zero MI which persists up to such large length
scales.\\The integrated stellar mass of a galaxy is thought to be the
most fundamental property which largely influences its structure and
evolution \cite{wouter}. It is often taken as an impression of its
formation history and governs its future evolution
\cite{wouter}. Galaxies assemble gas through both mergers and as well
as various secular processes \cite{sancisi,dave}. This gas upon
condensation, give rise to molecular clouds which in turn become
stellar nurseries. Hence, high stellar mass represent more evolved
systems. Furthermore, stellar mass of a galaxy correlate with other
galactic properties, such as star formation rate, metallicity and halo
mass \cite{brin,tremonti,baldry,lara,manucci,kravtsov}. Thus,
different scaling relations involving stellar mass coupled with the
galaxy stellar mass function (GSMF), are hypothesized to pose salient
constraints for the formation and evolutionary models of galaxies
\cite{schaye,pillepich}.

We investigate if the stellar mass which is linked to both morphology
and environment contributes to the non-zero MI between
morphology and environment. Although the stellar mass
function of galaxies is reported to be similar in
  clusters and field up to $z \sim 1.5$
  \cite{vulcani/2013,calvi/2013,ann/2014,van/2020} , it is not the case when
  the environment is parametrized by the local number density. Under
  these presumptions the stellar mass function is reported to vary
with environment \cite{balogh01,zehavi02}. These variations can be
partly explained by halo biasing \cite{sheth99,mo02}. However biasing
can not provide a complete understanding of the connection between
galaxy and dark matter halo. Besides, the stellar mass function of
galaxies is also known to be sensitive to their morphology
\cite{vulcani11}. Denser regions are preferentially inhabited by high
mass haloes which are again mostly populated by high stellar mass
galaxies with early-type morphology \cite{bamford}. There is a fair
possibility that the observed MI between morphology
and environment is a consequence of these relations. Bamford
\cite{bamford} studied the environmental dependence of morphology
using the Galaxy Zoo \cite{lintott08} data and found that morphology
exhibits very weak environmental trends once the stellar mass is
fixed. We would like to understand the role of stellar mass on the
observed environmental dependence of morphology using an information
theoretic perspective \cite{pandey}. The present manuscript aims to
investigate whether environmental dependence and morphology dependence
of stellar mass of a galaxy can explain the observed MI between environment and morphology \cite{pandey}.

It is also important to verify the observational findings with the
current models of galaxy formation and evolution. Semi analytic models
of galaxy formation and evolution (SAM) are prevailing tools for
understanding the hierarchical growth of dark matter haloes and
temporal evolution of galaxies within
\cite{cole,somerville,guo2011}. Distinguishable from empirical
abundance matching \cite{conroy,moster} or halo occupation
distribution models \cite{berlind}, SAMs utilize a forward-modeling
approach and are designed such that it include sufficient baryonic
physics required for understanding galaxy evolution, although at a
much simplified and macroscopic level \cite{mitchell}. This
macroscopic nature of SAM make them computationally inexpensive
compared to hydrodynamical simulations. However, predictions of SAMs
need to be confirmed by observations to understand their efficiency
and applicability. We analyze observational data from the $16^{th}$
data release (DR16) of the SDSS and compare the results against SAMs
implemented on the Millennium simulation.\\The paper is organized as
follows: In Section II we describe the methodology, in Section III we
delineate the method of analysis. In Section IV we describe the data
used in the paper, in Section V we discuss the results and in Section
VI we present our conclusion.\\ Throughout this work, we calculate the
comoving distances from redshifts using the flat $\Lambda$CDM model
with $\Omega_{m}=0.315$, $\Omega_{\Lambda}=0.685$ and $h=0.674$
\cite{planck18}.

\section{Methodology}
%%%%%%%%%%%%%%%%%%%%%%%%%%%%%%%%%%%%%%%%%%%%%%%%%%%
In this work we are interested in investigating whether only
  environmental and morphology dependence of stellar mass are
  inadequate in explaining the observed
  MI 
  between
  morphology and environment.
%In the present work,
For the purpose of analysis we use the method proposed in Pandey \& Sarkar
\cite{pandey}. A brief description of the method is as follows. We
construct a sample comprising $\Psi$ number of galaxies within a cubic
region. 
The galaxies belong to two distinct populations which we represent by "population A" and "population B".
%All the galaxies within the cube are either ellipticals or
%spirals and we exclude galaxies with uncertain morphology. 
We proceed
to divide the cube into regular 
%rectangular
square grids each having a volume
of $(d \hmpc )^{3}$. Thus the cube is now fractionated into many
smaller cubic voxels each of sidelength $d \hmpc$. Let $\Phi_{d}$ be the
number of the cubic voxels for a fixed grid size of $d$ resulted after
fractionating the bigger cube. The aim here is to count the number of
%spirals and ellipticals
population A and population B galaxies  in each
voxel. Let $(\psi_{A})_{i}$ and $(\psi_{B})_{i}$ be the number of
%elliptials and spirals
population A and population B galaxies  settling
in the $i^{th}$ voxel respectively. Therefore,
$\psi_{i}=(\psi_{A})_{i}+(\psi_{B})_{i}$ is the total number of
galaxies located in the $i^{th}$ voxel. Summing over all the
$\Phi_{d}$ voxels yields $\Sigma^{\Phi_{d}}_{i=1} (\psi_{A})_{i}=
\psi_{A}, \Sigma^{\Phi_{d}}_{i=1} (\psi_{B})_{i}=\psi_{B}$ and
$\Sigma^{\Phi_{d}}_{i=1} (\psi_{i})_{i}= \Psi$, where $\psi_{A}$,
$\psi_{B}$ and $\Psi=\psi_{A}+\psi_{B}$ are the total number of
population A, population B 
% ellipticals, spirals
and total number of galaxies in the sample respectively. We proceed to
define two discrete random variables $X$ and $Y$ with respective
probability distributions $P(X)$ and $P(Y)$, where $P(X_{i}) =
\frac{\psi_{i}}{\Psi}$ represents the probability of a galaxy chosen
at random settles in the $i^{th}$ voxel and $P(Y)$ is the probability
of a galaxy chosen at random is either a population
  A or a population B galaxy. Clearly $P(X)$ has $\Phi_{d}$ number
of possibilities while $P(Y)$ has only two and is given by $P(Y_{1}) =
\frac{\psi_{A}}{\Psi}$ for population A and
$P(Y_{2}) = \frac{\psi_{B}}{\Psi}$ for population
  B. We vary the grid size $d$ and compute $P(X)$ in each case. We
note that $P(Y)$ is insensitive to the grid size and remain constant
throughout the analysis. The total number of galaxies within each
voxel changes as the grid size $d$ is changed but the total number of
galaxies within the cube is constant. We compute a set of information
theoretic measures based on the counts in these voxels.

\subsection{The Mutual Information between Morphology and Environment}
In information theory, the average amount of information necessary to
explain a random variable $x$ is called the information entropy $H(x)$
and reads \cite{shannon}
\begin{eqnarray}
H(x) &= &-\sum_{i=1}^{n} p(x_{i}) \log p(x_{i})
\end{eqnarray}
where $p(x_{i})$ denote the probability of the $i^{th}$ outcome and
$n$ is the total number of outcomes. We set the base of the logarithm
to be $10$ throughout the analysis.

We have two variables $X$ and $Y$ defined for the environment on scale
$d$ and morphology respectively. We would like to measure the
association between these two random variables on different length
scales $d$. The joint entropy $H(X,Y)$ for the two random variables
$X$ and $Y$ in this case would be,
\begin{eqnarray}
H(X,Y)&=& - \sum_{i=1}^{\Phi_{d}} \sum_{j=1}^{2} P (X_{i},Y_{j}) \log P (X_{i},Y_{j})
\end{eqnarray}
where $P(X,Y)$ is the joint probability distribution for $X$ and
$Y$. The joint probability $P(X,Y)$ that a galaxy of a particular
morphology (population A or population B) chosen at random settles in the
$i^{th}$ voxel reads
\begin{eqnarray}
P(X,Y) &=& P(Y|X)P(X)
\end{eqnarray}
where $P(Y|X)$ is the conditional probability of a galaxy chosen at
random is either population A or population B  given that it settles in the
$i^{th}$ voxel.
Finally, the MI $I(X;Y)$ between $X$ and $Y$ is given
as \cite{bell62}
\begin{eqnarray}
I(X;Y)&=& \sum_{i=1}^{\Phi_{d}} \sum_{j=1}^{2} P (X_{i},Y_{j}) \log \frac{P (X_{i},Y_{j})}{P (X_{i})(Y_{j})}  \nonumber \\
&=&H(X) + H(Y) - H(X,Y)
\end{eqnarray} 
$I(X;Y)\geq 0$ as $H(X)+H(Y)\geq H(X,Y)$ with equality only when $X$
and $Y$ are independent. We compute $I(X;Y)$ on different length
scales by varying the grid size $d$.

The MI quantifies the expected gain in information on
  one variable given that the other one is observed. In the framework
  of information theory, this implies the reduction in uncertainty in
  estimating one variable given the knowledge of the other. A high
  value of $I(X;Y)$ indicates that both the random variables depend
  strongly on each other and therefore corresponds to a large
  reduction in uncertainty and vice versa. One particular advantage
  offered by MI is that it does not assume the nature of relationship
  between the two random variables and hence is sensitive to both
  linear and non-linear correlations.\\ It can be also expressed as a
  Kullback-Leibler (KL) divergence \cite{kullback} or relative entropy
  which is the difference between cross entropy and entropy. When the
  true distribution A is unknown, one may choose another distribution
  B as a model that approximates A. The cross-entropy is thus defined
  as $H(A,B)=-\sum_i\, A_i\, \log\,B_i$ and KL divergence is defined
  as, $KL(A||B)=\sum_i\,A_i\,\log\,\frac{A_i}{B_i}$. So the mutual
  information can be defined as the following KL divergence,
\begin{eqnarray}
\nonumber I(X;Y) & = & \sum_{i} \sum_{j} \, P(X_i,Y_j) \, \log\, \frac{P(X_i,Y_j)}{P(X_i)P(Y_j)} \\ & = & KL(P(X,Y)||P(X)P(Y))
\end{eqnarray}
Thus the mutual information can be also considered as the error of
using $P(X)P(Y)$ to model the joint probability $P(X,Y)$. When $X$ and
$Y$ are independent of each other i.e. $P(X,Y)=P(X)P(Y)$ then their
mutual information is zero.

\subsection{The Mutual Information between Stellar Mass and Environment}
We now define another random variable $Z$ which characterizes the
stellar mass of the galaxies. Similar to the previous section, the MI
$I(X;Z)$ between environment and stellar mass is defined as,
\begin{eqnarray}
I(X;Z)&=& \sum_{i=1}^{\Phi_{d}} \sum_{k=1}^{2} P (X_{i},Z_{k}) \log \frac{P (X_{i},Z_{k})}{P (X_{i})(Z_{k})} \nonumber \\
&=&H(X) + H(Z) - H(X,Z)
\end{eqnarray} 
where $k=1$ for low stellar mass galaxies and $k=2$ for high stellar
mass galaxies. In this work we treat galaxies having mass higher or
equal to the median stellar mass of the sample as being high mass
galaxies and vice versa. The joint entropy for environment and stellar
mass is given as
\begin{eqnarray}
H(X,Z)&=& - \sum_{i=1}^{\Phi_{d}} \sum_{k=1}^{2} P (X_{i},Z_{k}) \log P (X_{i},Z_{k})
\end{eqnarray}
\subsection{The Conditional Mutual Information between Environment, Morphology and Stellar Mass}
The positive MI between morphology and environment
  does not by definition mean a causal dependence between these
  galactic properties. The MI between environment and morphology may
come from the shared MI between stellar mass,
environment and morphology. To investigate this, we shall define the
conditional mutual information (hereafter CMI) between between stellar
mass, morphology and environment.

The CMI \cite{wyner78} furnish the expected value of the MI 
between two random variables given that we have complete knowledge of a 
third random variable. In this work, we are investigating the MI between morphology and environment of the galaxies given
the knowledge of their stellar masses.

The CMI between $X$ and $Y$ for a given $Z$ reads,
\begin{eqnarray}
I(X;Y|Z)&=&\sum_{i=1}^{\Phi_{d}} \sum_{j=1}^{2} \sum_{k=1}^{2} P (X_{i},Y_{j},Z_{k}) \log \frac{P (X_{i})P(X_{i},Y_{j},Z_{k})}{P (X_{i},Z_{k})P (Y_{j},Z_{k})} \nonumber \\
&=&H(X,Z) + H(Y,Z) - H(X,Y,Z) - H(Z)
\end{eqnarray}
where $H(X,Y,Z)$ is the joint entropy for the three random
variables. The joint entropy $H(X,Y,Z)$ is defined as,
\begin{eqnarray}
\nonumber H(X,Y,Z) & = & - \sum^{\Phi_{d}}_{i=1}
\sum^{2}_{j=1}\sum^{2}_{k=1} \, P(X_i,Y_j,Z_k) \, \log\,
P(X_i,Y_j,Z_k) \\
\label{eq:joint2}
\end{eqnarray}

The joint probability denoted by $P(X_i,Y_j,Z_k)=P(Z_k|Y_j,X_i)P(Y_j|X_i)P(X_i)$
can be estimated by counting the galaxies of different morphology and
stellar mass in $\Phi_d$ different voxels within the datacube. 

The CMI $I(X;Y|Z) \geq 0$ with equality only when the information
shared between $X$ and $Y$ is just a byproduct of the information
contained about these variables in $Z$. In other words, $I(X;Y|Z)= 0$
implies that $X$ is independent of $Y$ given the knowledge of $Z$.\\We
vary the grid size $d$ and compute $I(X;Y|Z)$ corresponding to each
length scale.

\section{Data}
\subsection{The SDSS sample}
We use data from the $16^{th}$ data release \cite{ahumada19} of Sloan
Digital sky survey (SDSS) for the present analysis. SDSS is presently
the largest redshift survey which has mapped more than one third of
the celestial sphere. The data is downloaded from the SDSS
SkyServer\footnote{https://skyserver.sdss.org/casjobs/} using
structured query language. The {\it SpecPhotoAll} and {\it Photoz}
tables of SDSS database are used to get the spectroscopic and
photometric information of the galaxies. We look for a region devoid
of any holes or patches so we select the contiguous region of the sky
within $0^{\circ} \leq \delta \leq 60^{\circ}$ and $ 135^{\circ} \leq
\alpha \leq 225^{\circ}$, $\alpha$ and $\delta$ respectively being the
right ascension and declination of a galaxy. We select the galaxies
within the $r$-band Petrosian apparent magnitude ($r_{p}$) limit
$r_{p} < 17.77$.  We obtain the morphology of the galaxies from the
{\it ZooSpec} table of DR16 database.  This table holds morphological
classification for SDSS galaxies provided through the Galaxy Zoo
project\footnote{http://zoo1.galaxyzoo.org}. In Galazy Zoo
\cite{lintott08, lintott11} millions of registered volunteers take
part in visual classification of the galaxies. We note
  that in the present analysis population A galaxies represent
  ellipticals and population B galaxies represent spirals. We
consider only the classified galaxies which are flagged as {\it
  spiral} or {\it elliptical} with a debiased vote fraction
$>0.8$. The galaxies which are flagged as {\it uncertain} are
discarded from the present analysis. We obtain the stellar mass of
galaxies from the {\it stellarMassFSPSGranWideNoDust} table. The
stellar mass of SDSS galaxies provided in this table are calculated
using the Flexible Stellar Population Synthesis (FSPS) techniques
\cite{conroy09}. Combining these four tables from the SDSS database,
we get a total $136108$ galaxies up to redshift $z < 0.3$. We use
these galaxies to prepare a volume limited sample with $r$-band
absolute magnitude limit $M_r \leq -21$ , stellar mass range $ 1\times
10^{10} M_{\odot} \leq M_{*} \leq 6 \times 10^{11} M_{\odot}$ and
redshift range $z \leq 0.115$. The median mass of the
  sample is $\simeq 8.054 \times 10^{11} M_{\odot}$. The resulting
volume and mass limited sample contains a total of
$43092$ galaxies. However, the present analysis requires a cubic
region. The largest cube that we are able to extract from the volume
limited sample has a side length of $174 \hmpc$. There are in total
$15373$ galaxies within the cube. $11481$ of them are spirals and
$3892$ are ellipticals.

\subsection{The Millennium samples}
\begin{figure*}[htbp!]
\centering
\includegraphics[width=12cm]{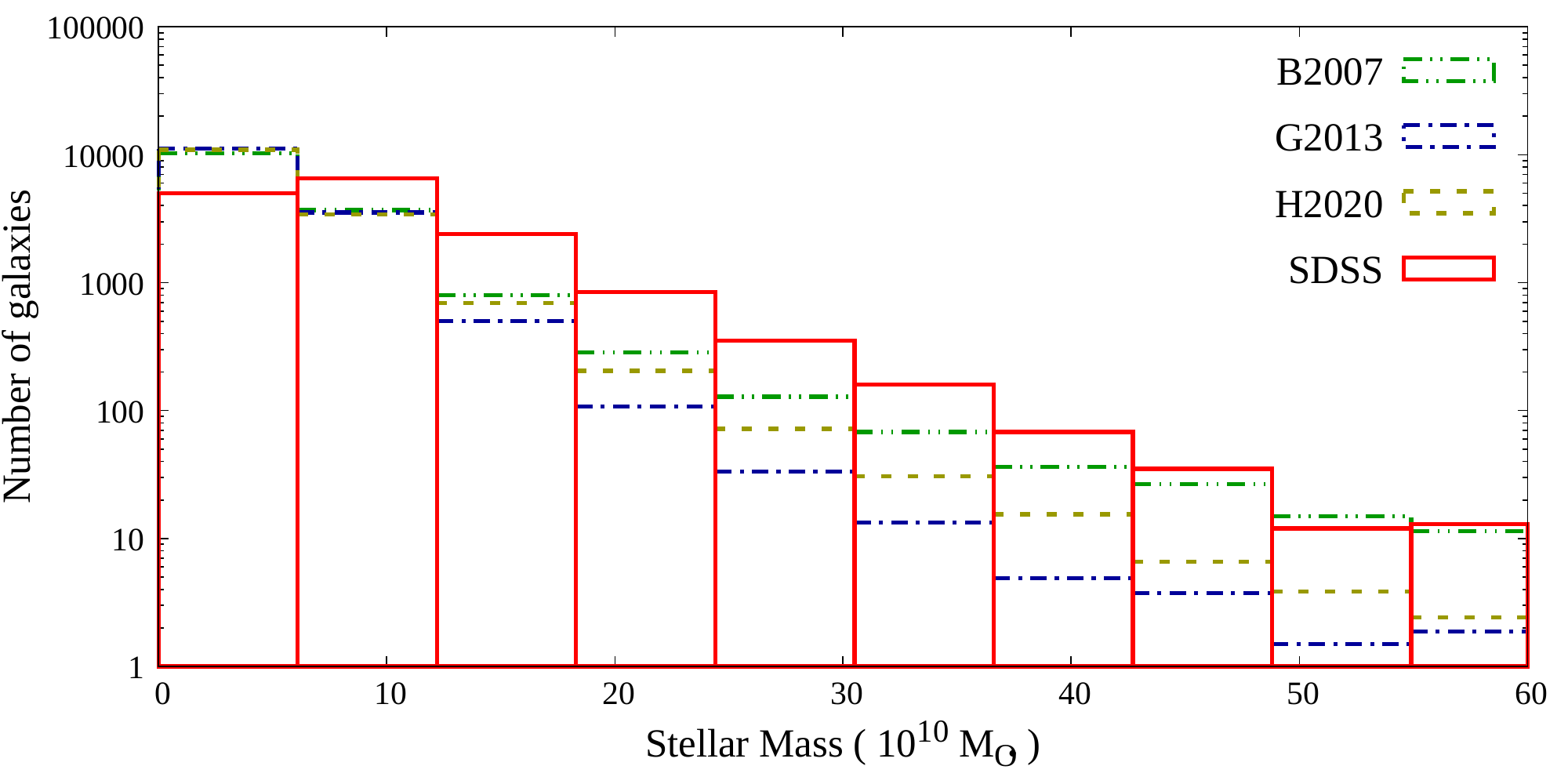}
\caption{This shows the number of galaxies as a
    function of stellar mass for the SDSS datacube and the mock
    datacubes from the SAMs. The number of
    galaxies in different stellar mass bins for the SAMs are averaged over 8 respective subcubes.}
\label{f2}
\end{figure*}

The Millennium Simulation is a cosmological simulation \footnote{
  http://www.mpa-garching.mpg.de/galform/virgo/millennium/} carried
out by the Virgo Consortium \footnote{ http://www.virgo.dur.ac.uk/ }
and described in detail in \cite{springel}. The simulation tracks
$2160^{3}$ particles each of mass $8.6 \times 10^{8} h^{-1} M_{\odot}$
enclosed in a comoving box of size $500 \hmpc$ on a side and assumes
$\Lambda$CDM cosmology with cosmological parameters $\Omega_{m}=0.25$,
$\Omega_{b}=0.045$, $h=0.73$, $\Omega_{\Lambda}=0.75$, $n=1$,
$\sigma_{8}=0.9$.  Since it is not practically
  feasible to analyze the large number of SAMs available in literature
  in one study, we shall therefore use the data of only 3 SAMs
  implemented on the Millennium Simulations. We analyze here for
  representative purposes the SAMs published by S. Bertone et al
  \cite{b2007} (hereafter B2007), Q. Guo et al \cite{g2013} (hereafter
  G2013) and by B.M.B Henriques et al \cite{h2020} (hereafter
  H2020). A brief description of the SAMs are as follows: The B2007
  model \cite{b2007} presents an enhanced feedback scheme constructed
  upon a dynamical treatment of the evolution of galactic winds
  replacing the previous use of supernova feedback without altering
  the AGN feedback and with recycling and ejection of gas and metals
  being treated self-consistently governed by the dynamical evolution
  of winds. The G2013 model \cite{g2013} investigates how the growth
  of structures vary with cosmological parameters derived from WMAP1
  \& WMAP7 by scaling the Millennium and Millennium II simulations
  using the re-scaling technique developed by Angulo \& White
  \cite{angulo/white}. Finally, the H2020 model \cite{h2020} is a
  recently published spatially resolved model of cold gas
  partitioning, star formation, mass and elements feedback and
  observationally consistent global properties which is argued to
  offer new opportunities to interpret the results on ongoing galaxy
  surveys.
% We obtain the data from three semi analytic galaxy
%formation models (SAM) by S. Bertone et al \cite{b2007} (hereafter
%B2007), by Q. Guo et al \cite{g2013} (hereafter G2013) and B.M.B
%Henriques et al \cite{h2020} (hereafter H2020). 
We download the data from the Millennium database
\footnote{http://gavo.mpa-garching.mpg.de/Millennium/} using a
Structured Query Language (SQL) search. We extract the position
coordinates, peculiar velocities, bulge mass and total stellar masses
of all the galaxies with $r$-band absolute magnitude limit $M_r \leq
-21$ and stellar mass range $ 1\times 10^{10} M_{\odot} \leq M_{*}
\leq 6 \times 10^{11} M_{\odot}$. We curve out $8$ non-overlapping
cubic regions of side $174 \hmpc$ from each of the SAM
catalogues. Note that the simulated galaxies are in real space whereas
the observed galaxies are in redshift space. Hence, for a fair
analysis, we map the 3D comoving coordinates of the simulated galaxies
in redshift space employing their peculiar velocities. As the
morphology of the galaxies are not directly available, we divide the
mass of the bulge $M_{bulge}$ to the stellar mass $M_{stellar}$ (i.e,
$\frac{M_{bulge}}{M_{stellar}} = \Theta$) for each galaxy and use this
parameter as a proxy for morphology. Since ellipticals are mostly
bulge dominated, we define all galaxies with $\Theta \geq 0.8$ as
ellipticals. Spirals are known to have a relatively smaller bulge
component. We define spirals as galaxies with $\Theta \leq 0.4$. We
discard all simulated galaxies with $ 0.4< \Theta <0.8$. This ensures
that the selected galaxies are primarily of two types - bulge
dominated and disk dominated. We randomly extract $15373$ galaxies
from each of these distributions in redshift space keeping the same
ratio of spirals to ellipticals. Finally for each SAM, we end up with
$8$ cubes with side length $174 \hmpc$.  Each of these datacube host
$11481$ spiral galaxies and $3892$ elliptical galaxies.
% The distribution of the spiral and elliptical galaxies in
%the SDSS datacube is shown in FIG. \ref{f1}. We also show the
%distribution of spirals and ellipticals in one mock datacube from each
%of the semi analytic models in FIG. \ref{f1}. 
We show the distribution
of stellar mass in the galaxy samples from the SAMs in
Fig. \ref{f2}. The result for the SDSS sample is also shown together
for a comparison.  The mock datacubes from the SAMs are analyzed
exactly the same way as the SDSS data.

%\begin{figure*}[htbp!]
%\centering
%\includegraphics[width=5.5cm]{sdss.pdf} 
%\includegraphics[width=5.5cm]{b2007.pdf} 
%\includegraphics[width=5.5cm]{g2013.pdf} 
%\caption{This shows the spatial distribution of spiral and elliptical 
%galaxies in the SDSS datacube. Distribution of spiral and elliptical 
%galaxies in one mock datacube from each of the two semi analytic models 
%are also shown separately. The blue dots represent the spirals and the 
%ellipticals are shown with red dots.}
%\label{f1}
%\end{figure*}

\section{Results}
\subsection{Significance of mutual information}
%%%%%%%%%%%%%%%%%%%%%%%%%%%%%%%%%%%%%%%%%%
In order to understand the connection between environment ($X$) and
morphology ($Y$), we calculate the MI $I(X;Y)$ between these
quantities for the SDSS sample.\\ We have restricted
  our analysis to the minimum length scale of $\sim 14 \hmpc$ as the
  mean intergalactic separation of our samples were of the order of
  $10 \hmpc$. A further decrease in the lengths of the voxels would
  result in an increase in the amplitude of fluctuations and therefore
  would yield spurious results. Additionally, note that the maximum
  length scale probed in this work is $87 \hmpc$ since the size of our
  data cube is $174 \hmpc$, therefore the maximum size of a voxel can
  only be $87 \hmpc$.\\ We find a positive MI between these
quantities which decreases with increasing length-scale similar to the
results obtained in \cite{pandey}. This indicates that environment do
influence morphology of any given galaxy and the degree of association
decreases on larger scales.  However, it is important to test the
physical significance of any non-zero MI observed between environment
and a galaxy property. To test the physical significance of the MI
between morphology and environment, we randomize the morphological
information of galaxies without affecting their spatial
distribution. We consider all the $15373$ galaxies in the SDSS
datacube and randomly tag $3892$ galaxies as ellipticals ignoring
their actual morphology. Rest of the galaxies in the SDSS datacube are
labelled as spirals. We repeat this procedure for $10$ times to
generate $10$ such SDSS datacubes with randomly assigned
morphology. This procedure randomize the information contained in the
morphology of galaxies keeping the galaxy distribution unchanged. The
adopted randomization procedure would not alter $H(X)$ and $H(Y)$ but
would only affect $H(X,Y)$. The value of $H(X,Y)$ should ideally tend
to $H(X)+H(Y)$ after randomization of morphology. Any physical
connection between morphology and environment is thus expected to be
destroyed by the randomization of morphological classification.

\begin{figure*}[htbp!]
\centering
\includegraphics[width=12cm]{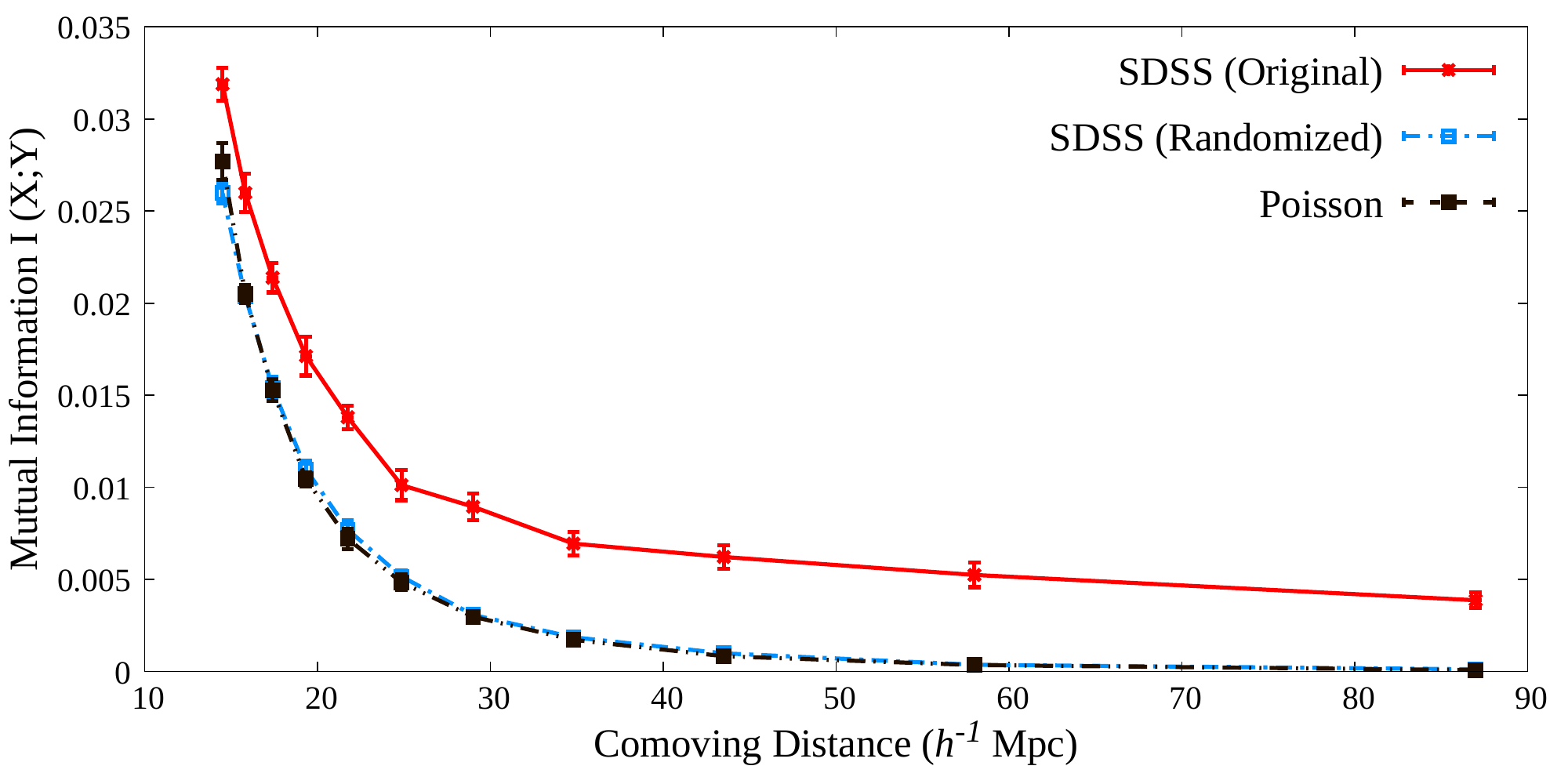}
%\captionsetup{width=.85\linewidth}
\caption{ This shows the MI $I(X;Y)$ between 
environment ($X$) and morphology ($Y$) for the original SDSS data and SDSS data with randomized 
morphological information. The error bars for the original SDSS data 
are obtained by bootstrap resampling whereas the error bars for the 
randomized dataset are obtained using $10$ such realizations.}
 \label{f3}
\end{figure*}

\begin{figure*}[htbp!]
%\begin{subfigure}{.52\textwidth}
\centering 
\includegraphics[width=7cm]{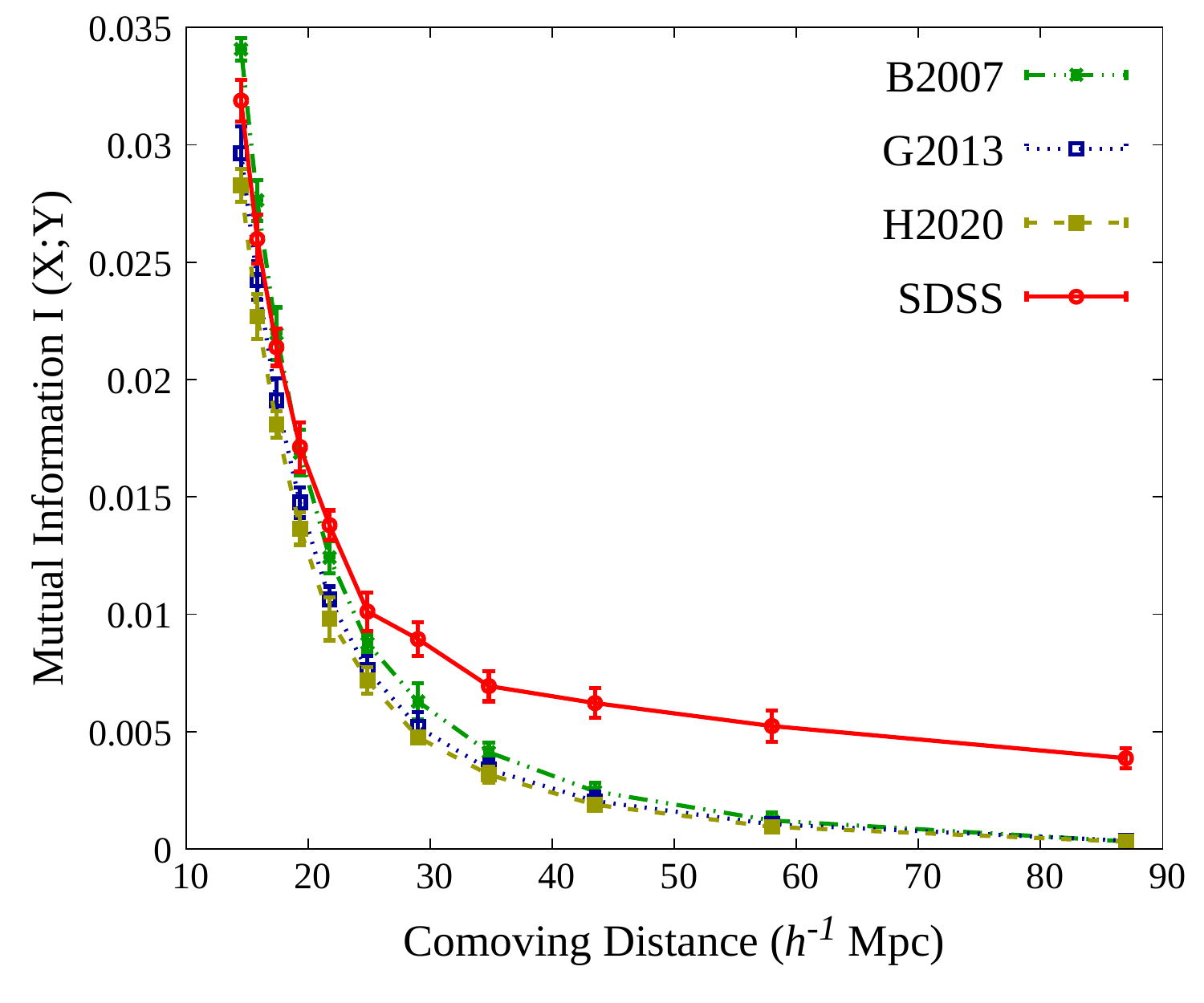} \hspace{0.25 cm}
\includegraphics[width=7cm]{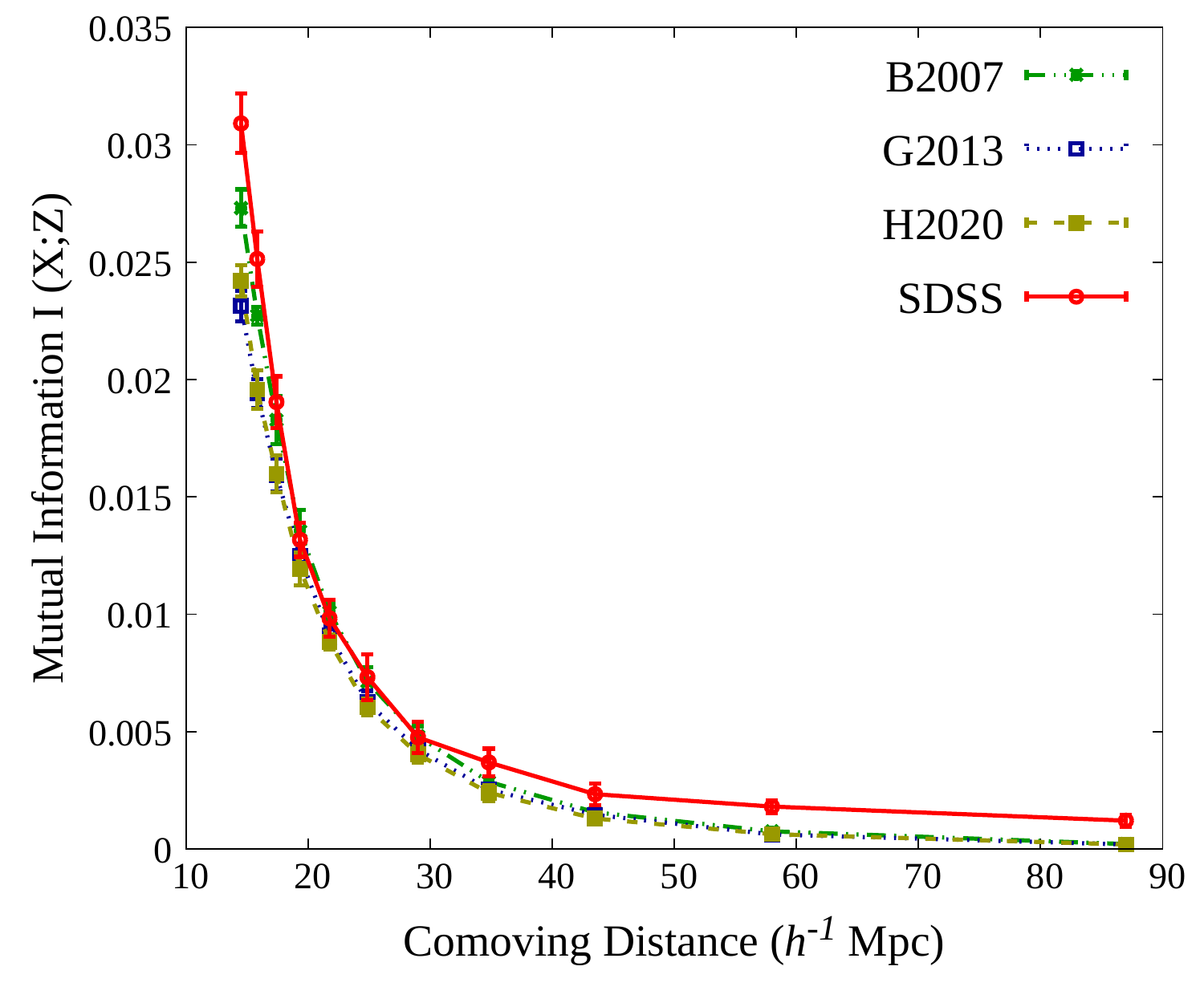}
% \captionsetup{width=.85\linewidth}
\caption{The left panel of this figure shows the MI $I(X;Y)$ between 
environment ($X$) and morphology ($Y$) as a function of length scales while the right panel shows the MI $I(X;Z)$
between environment ($X$) and stellar mass ($Z$) as a function of length scales for the SDSS and 
the semi analytical models. The 
error bars for the SDSS data are obtained by using $10$ bootstrap samples 
whereas the error bars for the semi analytic models are derived using $8$ 
subcubes drawn from each of the semi analytic catalogues.}
\label{f4}
\end{figure*}

The results for the SDSS data with original morphological
classification and its randomized counterparts are shown in
Fig. \ref{f3}. We find a significant reduction in the MI between morphology and environment when galaxies are
assigned a random morphology. However a smaller non-zero MI between morphology and environment still persists in the
randomized data which may originate from the finite size of the data
sample. We verify this by using a set of Poisson distributions. We
analyze $10$ mock Poisson datacubes each of which contain same number
of randomly assigned spirals and ellipticals as the SDSS datacube. We
find that the MI for the Poisson random distributions are nearly
identical to the SDSS data with randomized morphological
information. So the residual non-zero MI which survives after
randomization of morphological information of SDSS galaxies purely
arises due to the discrete and finite nature of the sample. A detailed
analysis of the statistical significance of these differences along
with some other tests in the present context are presented in a recent
paper \cite{sarkar}.

Thus a part of the MI between morphology and environment must have
some physical origin. The length-scale dependence of this MI may
contain important information regarding the physical processes
responsible for such correlations.

\begin{figure*}[htbp!]  
\centering
\includegraphics[width=12cm]{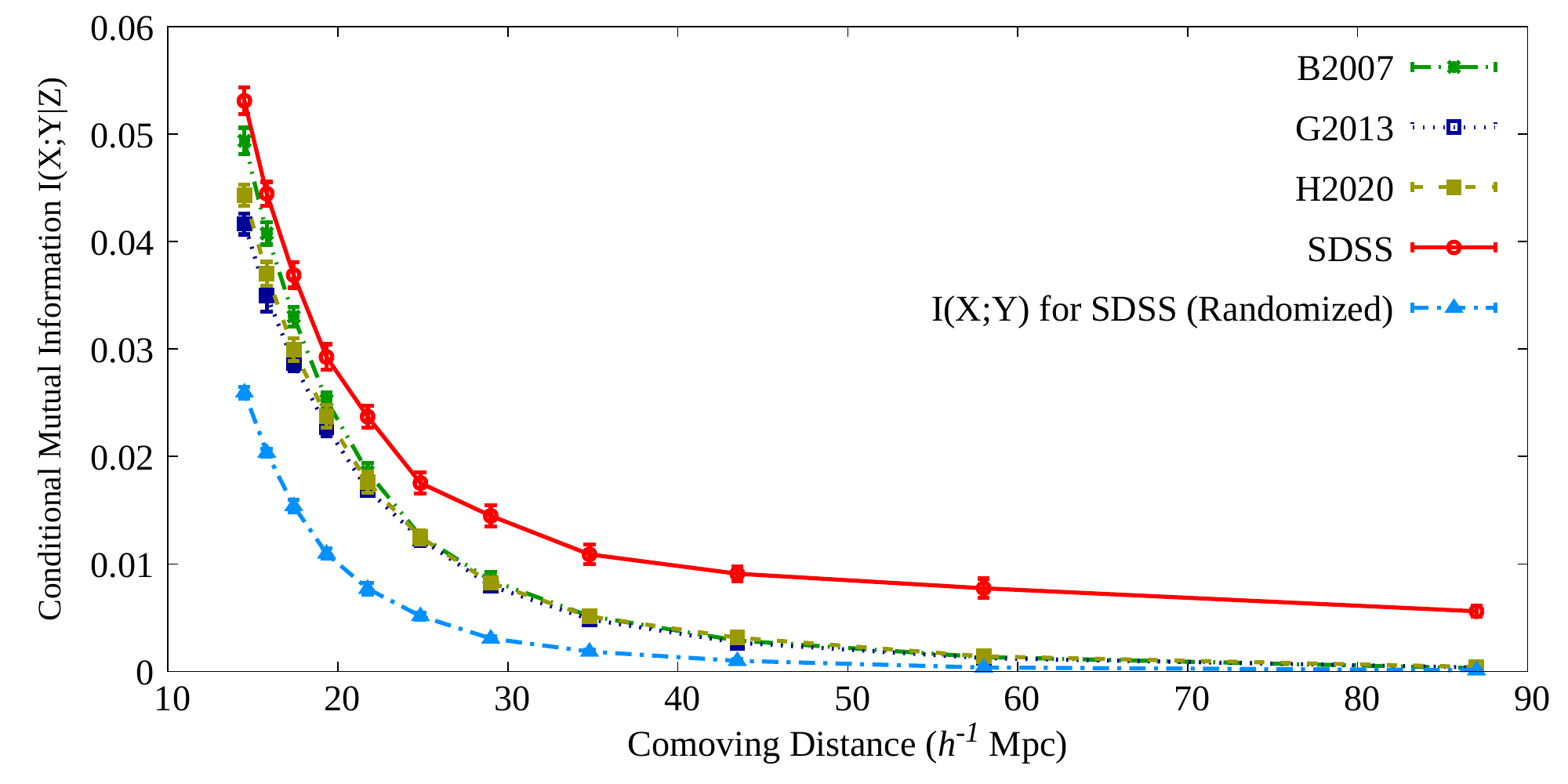}
%\captionsetup{width=.85\linewidth}
\caption{This shows the CMI $I(X;Y|Z)$ between 
environment ($X$), morphology ($Y$) and stellar mass ($Z$) as a function of length
  scales for the SDSS and the semi analytic models. The MI $I(X;Y)$
  between environment and morphology in the randomized SDSS data is
  shown together for a comparison. The error bars in
each case are obtained in the same way as Fig. \ref{f3} and
    \ref{f4}.}
\label{f5}
\end{figure*}

\subsection{MI and CMI: Comparing SDSS with SAM}
In this section we investigate whether the MI between morphology and
environment shown in Fig. \ref{f3} can be explained given a
knowledge of stellar mass. This possibility arises due to the fact
that stellar mass is correlated to both environment and morphology.

We first compare the MI between morphology and environment $I(X;Y)$ in
the SDSS with the SAMs considered. In the left panel of Fig. \ref{f4}, 
we show $I(X;Y)$ as a function of length scale for the SDSS and the SAMs. A
positive MI between environment and morphology is observed for both
the SDSS and the SAMs throughout the entire length-scale. The observed
MI between environment and morphology in the SAMs show a fairly
good agreement with SDSS on scales $<30 \hmpc$. However
the MI between morphology and environment in the SDSS exhibit higher
values compared to the SAMs on larger length scales ($>30 \hmpc$).

To carry out the analysis with stellar mass ($Z$), we have divided the
SDSS and SAM samples into two respective stellar mass bins. If the
stellar mass of a galaxy is $\leq 8.054 \times 10^{10} M_{\odot}$
(which is the median stellar mass of the SDSS sample) the galaxy
resides in the lower stellar mass bin and otherwise in higher stellar
mass bin. In the right panel of Fig. \ref{f4} we show the MI $I(X;Z)$
between environment ($X$) and stellar mass ($Z$) for the SDSS and
SAMs. We observe a positive MI between environment and stellar mass in
the SDSS as shown in the Fig. \ref{f4}. We find that the MI between
stellar mass and environment in the SAMs agree fairly well with
the SDSS observations. We observe some small differences in $I(X;Z)$
between SDSS and SAMs at smaller ($<20 \hmpc$) and larger ($>40
\hmpc$) length scales.

A positive MI between environment and morphology can be a byproduct of
the MI between environment and stellar mass. In other words, all the
information morphology shares with environment can be explained from
the knowledge of its stellar mass. To investigate this further, we
compute the CMI between environment and morphology by conditioning the
stellar mass (i.e, $I(X;Y|Z)$) in Fig. \ref{f5}. A perfect zero CMI
would suffice the argument that stellar mass indeed is capable of
explaining the MI between environment and
morphology. However a perfect zero MI is not expected even in a
randomized distribution or Poisson distribution where the two
variables are known to be uncorrelated.  We may recall that the MI
$I(X;Y)$ for the randomized SDSS data and the Poisson random
distribution are identical at all length scales. So the MI between
morphology and environment can be entirely explained by stellar mass
only if the CMI turns out to be positive and equal to the MI expected
for a Poisson random distribution.

The Fig. \ref{f5} shows that the CMI in SDSS is
significantly higher than the MI $I(X;Y)$ expected for the SDSS data
with randomized morphological information. This implies that
environmental and morphology dependence of stellar mass are not
sufficient in explaining the correlation between morphology and
environment at different length scales. The CMI predicted by the SAMs
are also much higher as compared to the MI for random
distributions. The CMI for both SDSS and SAMs are positive and larger
than the non-zero MI expected from random distributions at all length
scales which indicates that a part of the MI between morphology and
environment may arise due to the associations of stellar mass with
environment and morphology. However these correlations are unlikely to
explain the entire MI between morphology and environment. We note that
B2007 model predict a relatively higher MI and CMI than the G2013 and
H2020 model on length scales $<30 \hmpc$. The three SAMs predict
nearly the same MI and CMI on length scales $>30 \hmpc$. We find that
the CMI measured in the SDSS is larger than the CMI predicted by the
SAMs at all length scales. The results also indicate the existence of
complex physical processes that alters the stellar mass or star
formation of galaxies without affecting their morphologies and
vice-versa. However, there is no general consensus as to what these
processes are and therefore, future observations coupled with
state-of-the art simulations should focus on answering this
issue. \\ It may be noted that the Mutual Information
  $I(Y;Z)$ between morphology $(Y)$ and stellar mass $(Z)$ is not
  explored since $I(Y;Z)$ is independent of length scale and therefore
  do not aid significantly in our investigation to understand how the
  correlations between two or more galactic properties vary with
  length scale.
\section{Conclusions}
In this paper we investigate the possibility that a prior knowledge of
stellar mass of galaxies may explain the MI between
environment and morphology. To investigate this, we employ an
information theoretic framework developed in \cite{pandey} to the
publicly available SDSS DR16 and compare the results against semi
analytic models implemented on the Millennium simulation.

We find a non zero MI between environment and morphology $I(X;Y)$ for
both SDSS and semi analytic models which indicates that morphology and
environment are correlated up to several tens of Mpc. The MI between
morphology and environment decreases with increasing length scales but
remain non-zero throughout. We show that randomizing the information
about the morphology of SDSS galaxies leads to a significant reduction
in the MI between environment and morphology. This indicates that the
randomization destroys the physical association between morphology and
environment. We find a residual non-zero MI which is equal to that
expected for a mock Poisson random distribution. This residual part of
the non-zero MI originates from the finite and discrete nature of the
distribution. However the excess part of the non-zero MI between
morphology and environment owes an explanation. We investigate if this
additional non-zero MI is a consequence of the fact that stellar mass
is separately related to both morphology and environment.

We find a non-zero CMI between morphology, environment and stellar
mass in the SDSS throughout the entire length scales probed. The
observed CMI is significantly larger than the MI between morphology
and environment expected in a randomized distribution with no existing
correlation between the two. We conclude from the excess non-zero CMI
that the environmental and morphological trends of stellar mass are
inadequate in explaining the MI between morphology and
environment. We compare these findings with three different
semi-analytic models all of which predict significantly lower MI and
CMI compared to SDSS on larger length-scales. It is
  worthwhile to mention here that our analysis pertains to the local
  Universe only.\\

All three semi-analytic models (B2007,G2013, H2020)
  analyzed in this work, adopted a Chabrier initial mass function
  \citep{chabrier} in their modelling. The primary differences in the
  semi-analytic models lie in their choice of the cosmological
  parameters and the schemes for modelling various physical processes
  responsible for galaxy formation and evolution.  The feedback
  processes (e.g. SNe feedback, AGN feedback) can significantly affect
  the growth of galaxies and shape their mass functions. Expulsion of
  gas from galaxies by SNe feedback drive the evolution of low mass
  galaxies, but do not significantly affect the evolution of high mass
  galaxies. The AGN feedbacks which are driven by the super-massive
  black holes (SMBH) at the centre of galaxies, suppress cooling in
  massive halos leading to a quenching in star formation. This turns
  the massive galaxies into ``red and dead" ones which subsequently
  grow by mergers and become bulge dominated in the process. The
  choice of feedback schemes can thus indirectly affect the
  morphological mix of galaxies in different SAMs.  The B2007 model
  used a new feedback scheme with a more efficient AGN feedback than
  those implemented in G2013 and H2020 models. The higher efficiency
  of the AGN feedback in the B2007 model is expected to produce a
  larger number of galaxies with higher stellar mass. This can be
  clearly seen in Fig. \ref{f2}. The
  Fig. \ref{f2} also shows that the abundance of low
  stellar mass galaxies are nearly same in all the SAMs.

It is interesting to note that the MI and CMI (shown
  in Fig. \ref{f4} and
  Fig. \ref{f5}) in the B2007 model show a somewhat
  better agreement with observations than the other two models on
  scales below $30 \hmpc$. A more efficient AGN feedback in B2007 model
  yields a larger number of high stellar mass galaxies. The
  small-scale environments in the cosmic web may play an important
  role in the growth of the SMBHs in these galaxies \citep{umehata}
  which are responsible for triggering AGN activities. The corelation
  between the environment and growth of SMBH may thus indirectly
  influence the mutual information between environment and galaxy
  properties such as morphology and stellar mass.

The MIs and CMIs in all the models are nearly
  identical on scales beyond $30 \hmpc$. This indicates that the
  differences in the modelling of different physical processes and the
  associated parameters have hardly any influence on the measured MI
  and CMI on these length scales. The assembly bias \citep{croton,
    gao, musso, vakili} in simulations show that the early-forming
  low mass haloes are strongly clustered than the late-forming haloes of
  similar mass, which arises due to the differences in the accretion
  and merger histories of the haloes across different environments. The
  large-scale correlations observed both in SDSS and the models may be
  a consequence of the assembly bias which requires further studies.
  However, the discrepancy between the observed MI and CMI in SDSS and
  those predicted by different SAMs on scales beyond $30 \hmpc$
  indicates that a better understanding of the role of large-scale
  environment is needed to account these differences.

Finally we note that morphology of galaxies are indeed correlated to
their environment up to large length scales and we still lack a
complete understanding of the physical origin of such large-scale
correlations.

\section*{Acknowledgments}
%%%%%%%%%%%%%%%%%%%%%%%%%%%%%%%%%%%%%%%
  The authors thank an anonymous reviewer for useful comments and
  suggestions which helped us to improve the draft. The authors would
  like to thank the SDSS team and the Galaxy Zoo team for making the
  data public. The Millennium Simulation databases \cite{lemson} used
  in this paper and the web application providing online access to
  them were constructed as part of the activities of the German
  Astrophysical Virtual Observatory (GAVO). BP acknowledge Steven P. Bamford for useful discussions. SB thanks Department of
  Science and Technology, New-Delhi, Government of India for the
  provisional INSPIRE fellowship selection
  [No:DST/INSPIRE/03/2019/003141]. BP would like to acknowledge financial
  support from the SERB, DST, Government of India through the project
  CRG/2019/001110. BP would also like to acknowledge IUCAA, Pune for
  providing support through associateship programme. SS would like to
  thank UGC, Government of India for providing financial support
  through a Rajiv Gandhi National Fellowship. SB and SS also thank Biswajit Das for discussions. 

\end{document}